\documentclass[aps,showpacs,11pt,letterpaper,nofootinbib,amsmath,amssymb]{revtex4-2}%
\usepackage{latexsym}%
\usepackage{epsf}%
\usepackage{graphicx}%
\usepackage{epsfig}%
\usepackage{graphicx}%
\usepackage{epstopdf}%
\usepackage[dvips,dvipdfmx,hypertex]{hyperref}%

\def\cO{{\mathcal{O}}}
\def\cM{{\mathcal{M}}}

\def\hh{{\widehat h}}

\def\cH{{\mathcal{H}}}

\def\cE{{\mathcal E}}

\DeclareMathAlphabet{\mathpzc}{OT1}{pzc}{m}{it}

\newcommand{\beq}{\begin{equation}}
\newcommand{\beqn}{\begin{equation}\nonumber}
\newcommand{\eeq}{\end{equation}}
\newcommand{\bea}{\begin{eqnarray}}
\newcommand{\bean}{\begin{eqnarray}\nonumber}
\newcommand{\eea}{\end{eqnarray}}
\newcommand{\ba}{\begin{align}}
\newcommand{\ea}{\end{align}}

\begin{document}

\title{A quantum thin shell surrounding a Black Hole}
\author{Cenalo Vaz\footnote{\tt Cenalo.Vaz@uc.edu}}
\affiliation{University of Cincinnati Blue Ash, Cincinnati, Ohio 45236.}
\begin{abstract}
 
In a previous work we obtained exact solutions for the proper time quantum mechanics of a thin dust shell, collapsing in a vacuum. We 
extend these results to the quantum collapse of a dust shell surrounding a pre-existing black hole.  In lieu of exact solutions, which 
have so far proved difficult to obtain for this system, we establish the essential features of the quantum shell through a 
Wentzel–Kramers–Brillouin (WKB) approximation, which is valid only when the mass of the shell is much greater than the Planck mass. 
There are many similarities with the vacuum collapse: only bound states exist and the proper energy spectrum of the shell is unaffected 
by the presence of the central black hole to this order. There are no peculiar or distinguishing features of the wave function near the 
black hole horizon. It vanishes at the center and oscillates between the origin and the classically forbidden region, beyond which it 
decays exponentially.
\pacs{04.60.Ds, 04.60.-m}
\end{abstract}
\maketitle

\section{Introduction}

Thin shells of matter, collapsing in a variety of environments, have been used extensively as simplified systems with which to model the final 
stages of gravitational collapse under many different conditions \cite{visser04,xu06,vachaspati08,vachaspati09,paranjape09,bardeen14,stojkovic14,
ziprick16,binetruy18,baccetti20}. This is because thin shell collapse captures many of the features of more realistic collapse models while 
avoiding some of their technical difficulties. Moreover, the quantum mechanics of the shell is exactly solvable in some cases, which helps to
shine a light on some of the problems of quantum gravity. For example, ``time'' has different meanings in classical general relativity and in the 
quantum theory. All choices of the time function yield the same local geometries, but quantum theories built on different time parameters are 
not unitarily equivalent. In \cite{vaz22}, we showed that exact quantizations, based on different time variables, of a shell that is 
collapsing in a vacuum yield incompatible descrptions. When the shell quantization is based on coordinate time, solutions exist only when 
its mass is {\it less} than the Planck mass \cite{hajicek92a}, but when it is based on proper time, solutions exist only when its mass is 
{\it greater} than the Planck mass, which is more in keeping with what is observed. 

Among other important issues that one would like to understand from the point of view of quantum gravity is the Hawking effect \cite{hawking75}
and the information loss paradox in black hole physics. Most discussions of the Hawking effect examine particle production in a scalar field 
propagating in the classical background geometry of a collapsing body from the point of view of the asymptotic observer. This is an effective field 
theory approach from which black hole evaporation and information loss are inferred. The geometry excites scalar field quanta, which then propagate 
to infinity as thermal, nearly thermal or unitary radiation (recently, it was argued in \cite{stojkovic15,stojkovic16} from this point of view 
that the radiation from a thin shell during its collapse is unitary). The result is that the black hole appears to evaporate over time as energy 
is drawn from it by the excited field quanta. Reasoning that this effect should have a counterpart in a time dependent quantum gravitational collapse 
and from the point of view of a comoving observer (the black hole either evaporates or it does not), we constructed a midisuperspace quantization 
of a non-rotating dust ball (the simplest form of collapse) \cite{vaz01,kiefer06} using the equivalent of Kucha\v r variables \cite{kuchar94}
in the LeMa\^itre-Tolman-Bondi (LTB) frame \cite{ltb}. We were able to build exact diffeomorphism invariant states on a lattice, thereby treating the dust ball as a series 
of shells labeled by their LTB radial coordinate, and showed that matching the shell wave functions across the apparent horizon requires ingoing 
modes in the exterior to be matched to outgoing modes in the interior and, vice versa, ingoing modes in the interior are matched to outgoing 
modes in the exterior \cite{vaz10,vaz13}. In each case, the relative amplitude of the outgoing wave is suppressed by the square root of the 
Boltzman factor at the Hawking temperature determined by the total Misner-Sharp mass contained within the shell. There are two independent 
solutions. In one, exterior, infalling waves representing the collapsing shells of dust are accompanied by interior, outgoing waves. These interior 
waves, which are of quantum origin, represent an interior ``Unruh'' radiation. In the other solution, waves move away from the apparent horizon 
on both sides of it. Interior, infalling waves representing the continued collapse of the dust shells across the apparent horizon are accompanied 
by exterior, outgoing waves. These latter ougoing waves represent the exterior Unruh radiation, which is thermal. Continued collapse across the 
apparent horizon from an initial diffuse state can be achieved by combining the two solutions and requiring the net flux to vanish at the apparent 
horizon. The effect is that the collapse ends in a central singularity and is accompanied by thermal Unruh radiation in the exterior. The 
net effect in the proper time quantum theory is therefore a quasi-classical tunneling of particles as described in \cite{parwil00,qqj06}, 
but only if a continued collapse beyond the apparent horizon is assumed.

There are some ambiguities involved in the midisuperspace quantization program that are avoided in the quantization of a thin shell. 
Therefore, in this paper, we address the quantum mechanics of single thin dust shell that is collapsing in the background of a pre-existing 
black hole from the comoving observer's point of view. The purpose is to understand what differences the background brings about in the 
shell wave function, in particular in its behavior at the event horizon of the black hole and at infinity, {\it i.e.,} is there a similar 
tunneling effect as described above for the dust ball and if so does the shell evaporate thermally as suggested by Hawking? 

On the classical 
level, the shell has just one degree of freedom and is completely described by its radius, $R(t)$ and its conjugate momentum, $P(t)$. We are 
unable to find exact wave functions, as we did in the case of a shell collapsing in a vacuum, but a WKB approximation is sufficient to extract 
many of their key features. We find several differences between the behavior of shells in a dust ball and the single shell, all of them traceable 
to the fact that, unlike the shells of a dust ball, the single shell possesses a self-interaction that is inversely proportional to its area 
radius. As a consequence the single shell is classically bound. We will show that there is no tunneling across the horizon and determine the 
energy spectrum of the shell. Its WKB wave function extends from the origin (where it vanishes), is well behaved at the black hole horizon and 
falls off exponentially beyond the classically allowed region.

In Sec. II we derive the proper time dynamics of a classical thin shell collapsing onto a pre-existing mass. The classical dynamics are 
obtained by an application of the Israel-Darmois-Lanczos (IDL) formalism \cite{israel66,darmois27,lanczos24}, which yields a first integral of 
the motion involving the pre-existing mass at the center, $M^-$, the Arnowitt-Deser-Misner (ADM) mass at infinity, $M^+$, and the proper mass, 
$m$, of the shell. Of these, the mass at the center and the proper mass of the shell are non-dynamical parameters, constant over the entire 
phase space. The ADM mass is a dynamical variable that represents the energy of the system. There are three time variables, {\it viz.,} the 
coordinate times in the interior and the exterior, and the shell proper time. What is not clear is the time variable in which the ADM mass 
generates the evolution. We follow Haji\v cek, Kay and Kucha\v r \cite{hajicek92a} and assume that it generates the evolution in the time 
coordinate inside the shell. This allows us to construct Lagrangians and Hamiltonian evolutions for the shell in the other time 
variables, in particular in the shell proper time. Here we also show that the Hamiltonian obtained in this way is structurally similar to the 
proper time Hamiltonian derived for the full Einstein-dust system in the LTB dust ball models, which lends confidence in 
the choice of \cite{hajicek92a}. In section III, we quantize the classical model. The Wheeler-DeWitt equation 
is elliptic with a positive semi-definite inner product for energies less than the shell's proper mass. We find the WKB approximation of the 
wave function and, in Sec. IV, analyze its $U(1)$ current. Requiring that a lowest energy state exists and that the $U(1)$ current is finite 
and well behaved everywhere, a complete set of bound states exists and we determine its spectrum. By comparing the solutions with the shell 
collapsing in a vacuum, we conclude that the approximation is valid only so long as the proper mass of the shell is much greater than the 
Planck mass. We conclude in Sec. V with a brief summary of our results and tie them in with a previously suggested model for quantum black 
holes.

\section{The Classical Shell Model}

The equation of motion of a spherical, thin, massive shell is obtained by applying the Israel-Darmois-Lanczos conditions on the timelike 
surface $\Sigma = \mathbb{R}\times \mathbb{S}^2$ that represents its world sheet. The world sheet forms the three dimensional boundary 
between an internal spacetime, $\cM^-$, and an external spacetime, $\cM^+$. $\cM^{\mp}$ are described in coordinates $x_\mp^\mu$ by metrics 
$g_{\mu\nu}^\mp$ that solve Einstein's equations. Let $\xi^a$ be a set of intrinsic coordinates on the surface of the shell and differentiable 
functions of $x^\mu_\mp$, then ${e^\mp}^\mu_a = \partial x_\mp^\mu/\partial \xi^a$ are the components of the three basis vectors on this 
surface and $h^\mp_{ab} = g^{\mp}_{\mu\nu} {e^\mp}^\mu_a{e^\mp}^\nu_b$ is the induced metric on the shell on the two sides of it. The first 
junction condition requires the shell to have a well defined metric, {\it i.e.,} $h^-_{ab} = h^+_{ab}$ or $[h_{ab}]=0$.

The second junction condition, which follows from Einstein's equations, says that the surface stress energy tensor, $S_{ab}$, of the shell 
is given by 
\beq
S_{ab} = -\frac\varepsilon{8\pi}\left([K_{ab}] - [K] h_{ab}\right)
\label{shellstress}
\eeq
where $K_{ab}$ is the extrinsic curvature of the boundary, $K = K^a_a$ and $\varepsilon = +1$ for a timelike shell.
If $\cM^\mp$ are taken to be vacuum spacetimes, then spherical symmetry implies that $g_{\mu\nu}^\mp$ are Schwarzschild metrics,
with mass parameters $M^\mp$ respectively, and $M^+$ represents the total mass of the system. We may write the respective line
elements as
\beq
ds^2_\mp = - g^\mp_{\mu\nu} d x_\mp^\mu d x_\mp^\nu = B^\mp dt_\mp^2 - \frac 1{B^\mp} dr^2_\mp - r_\mp^2 d\Omega^2
\eeq
where $B^\mp = 1-2GM^\mp/r_\mp$ and we have assumed that the interior and exterior share the same spherical coordinates, $\theta$ 
and $\phi$. The shell is described by the parametric equations $r_\mp = r = R(\tau)$, $t_\mp = t_\mp(\tau)$, where $\tau$ is the 
proper time for comoving observers and the interior and exterior time coordinates are related to the shell proper time (and
indirectly to each other) by
\beq
\frac{dt_\mp}{d\tau} = \frac{\sqrt{B^\mp + R_\tau^2}}{B^\mp},
\label{timeders}
\eeq
where the subscript indicates a derivative with respect to $\tau$. Choosing the intrinsic coordinates of the shell to be $\xi^a =
\{\tau,\theta,\phi\}$, the induced metric is
\beq
ds_\Sigma^2 = d\tau^2 - R^2(\tau) d\Omega^2.
\eeq
while the non-vanishing components of the extrinsic curvature are 
\beq
{K_\mp^\theta}_\theta = {K_\mp^\phi}_\phi = \frac{\beta^\mp}R,~~ {K_\mp^\tau}_\tau = \frac{\beta_\tau^\mp}{\dot R_\tau},
\eeq
where
\beq
\beta^\mp = \sqrt{B^\mp + \dot R_\tau^2}.
\eeq
Therefore, according to \eqref{shellstress}, 
\bea
{S^\tau}_\tau &=& \frac{\beta^+-\beta^-}{4\pi GR} = -\sigma\cr\cr
{S^\theta}_\theta = {S^\phi}_\phi &=& \frac{\beta^+-\beta^-}{8\pi GR} + \frac{\beta_\tau^+ - \beta_\tau^-}{8\pi G R_\tau} 
= p
\label{shellstress2}
\eea
where we have set ${S^a}_b =\text{diag}(-\sigma,p,p)$. 

The mass density of the shell is ``$\sigma$'' and ``$p$'' is its tangential pressure, which, for dust shells, we take to be zero. 
Integrating the second equation in \eqref{shellstress2},
\beq
\beta^+-\beta^- = -\frac {Gm}R,
\label{deltabeta}
\eeq
where $m$ is a constant of the integration, which represents the rest mass of the shell, as is seen by inserting this solution 
into the first. Equation \eqref{deltabeta} may be put in the form
\beq
M^+-M^- = \Delta M = m\sqrt{B^-+R_\tau^2} -\frac{Gm^2}{2R}.
\label{genconst}
\eeq
It is reasonable think of the above as a first integral of the motion and associate $\Delta M$ with the total energy, $E$,
of the shell. When expressed in terms of the momentum conjugate to $R(\tau)$, \eqref{genconst} will represent the Hamiltonian 
of the system. It is, however, given in terms of the velocities, which are dependent variables in the canonical theory and,
to determine the momentum, it becomes necessary to know in which of the three time coordinates the Hamiltonian is evolving 
the system. 

Within the thin shell construction, there is no {\it \`a priori} way to determine a canonical Hamiltonian 
because the constraint equation has been derived from the IDL conditions and not a fundamental action principle. 
One approach would be to compare the thin shell Hamiltonians with a similar system for which a canonical theory has been 
derived from an action principle. Our goal will be to recover a proper time Hamiltonian that is compatible with the 
midi-superspace Hamiltonian \cite{vaz01,kiefer06} obtained for the spherically symmetric Einstein-dust action by an application 
of a canonical chart analogous to that employed by Kucha\v r in \cite{kuchar94,brown94}. Because a dust ball can be thought of 
as a sequence of non-interacting shells, the proper time Hamiltonian for a single shell should be of the same form apart from 
any self-interaction terms peculiar to the thin shell itself. Thus, for example, if the evolution is taken to be in the shell 
proper time and the energy is taken to be $\Delta M$, we have
\beq
R_\tau = \frac{\partial H_I}{\partial p}
\eeq
and the Hamiltonian is \cite{berezin88,berezin97}
\beq
H_I = m \sqrt{B} \cosh \frac pm - \frac{Gm^2}{2R}.
\label{hyperbolicham}
\eeq
It does not have the same form as the Hamiltonian derived for the dust ball in \cite{vaz01,kiefer06}. 

As mentioned in the introduction, we will show that the choice of \cite{hajicek92a}, taking $\Delta M$ to evolve the system 
in the coordinate time of the interior, yields a compatible proper time Hamiltonian. Because the right hand side of \eqref{genconst} 
involves only the interior we drop the superscripts $\pm$ and, employing \eqref{timeders}, we can rewrite it as
\beq
\Delta M = \frac{mB^{3/2}}{\sqrt{B^2-R_t^2}} - \frac{Gm^2}{2R}.
\eeq
Then $R_t=\partial H_{II}/\partial p$ gives
\beq
H_{II} = - P_{(t)} = \sqrt{m^2B+B^2 p^2} - \frac{Gm^2}{2R},
\label{tham}
\eeq
and
\beq
p = \frac{m R_t}{\sqrt{B}\sqrt{B^2-R_t^2}}.
\label{momt}
\eeq
The action for the shell may now be given as a Legendre transform of $H_{II}$,
\beq
S = \int dt \left[-m \sqrt{B-\frac{R_t^2}B} + \frac{Gm^2}{2R}\right]
\eeq
and then transformed into an action in proper time, once again with the help of \eqref{timeders}. One finds
\beq
S = \int d\tau \left[-m + \frac{Gm^2}{2R}\frac{\sqrt{B+R_\tau^2}}B\right]
\eeq
and the proper time Hamiltonian
\beq
\cH = - P_{(\tau)} = m - \sqrt{\frac{f^2}B - BP^2},
\label{ptham}
\eeq
where we have set $f(R) = Gm^2/2R$ and the momentum, $P$, conjugate to $R$, is now given by 
\beq
P = \frac{fR_\tau}{B\sqrt{B+R_\tau^2}}.
\label{momtau}
\eeq
This proper time Hamiltonian is bounded from above by the mass of the shell and the shell momentum is bounded from above 
by $f/B$. From the comoving observer's point of view, the shell is always bound to the center.

The equations of motion that follow from \eqref{ptham} are derivable from the super-Hamiltonian
\beq
h_{(\tau)} = (P_{(\tau)} +m)^2 + B P^2 - \frac{f^2}B = 0.
\label{shtau}
\eeq
In this form, the Hamiltonian structure of the shell is identical to that of the dust ball, as derived in the Einstein-dust 
system, with one important exception: the shells in a dust ball do not possess a self-interaction that depends on their area 
radius: for the shells in a dust ball, $f(R)$ gets replaced by the Misner-Sharp mass density, $F'(r)$, where $r$ is the LTB shell 
label coordinate and $F(r)$ represents the mass of the dust ball up to $r$. The dependence of the self-interaction term, $f(R)$, 
on the area radius is responsible for the fact that the shell is classically bound in the proper time description. It will also 
play an important role in the matching conditions at the horizon.

\section{The Quantum Shell}

The structure of the super-Hamiltonian in \eqref{shtau} indicates that the DeWitt metric is 
\beq
\gamma_{ab} = \left(\begin{matrix}
1 & 0\cr
0 & 1/B \end{matrix}\right),
\eeq
so we choose a factor ordering that is symmetric with respect to the measure ``$dR/\sqrt{B}$''. Raising the momenta to operator 
status following Dirac we get the Wheeler-DeWitt equation
\beq
\hh_{(\tau)}\Psi(\tau,R) = \left[\left(-i\hbar\frac{\partial}{\partial\tau} +m\right)^2 - \hbar^2\sqrt{B}\frac{\partial}{\partial_R} \sqrt{B} 
\frac{\partial}{\partial_R} - \frac{f^2}{B}\right]\Psi(\tau,R) = 0.
\label{wd}
\eeq
We have been unable to find exact solutions to this equation, but the WKB approximation suffices to yield a general picture of the 
quantum shell. With $\Psi(\tau,R) = e^{iW(\tau,R)/\hbar}$, \eqref{wd} reads
\beq
-i\hbar \frac{\partial^2 W}{\partial\tau^2} + \left(\frac{\partial W}{\partial \tau} + m\right)^2 - i\hbar B \frac{\partial^2 W}
{\partial R^2} - \frac{i\hbar }2 B' \frac{\partial W}{\partial R} + B \left(\frac{\partial W}{\partial R}\right)^2 - \frac{f^2}B,
= 0
\eeq
where the prime indicates a derivative with respect to $R$, and taking
\beq
W = - \cE\tau + S_0(R) + \frac \hbar i \ln A(R),
\label{Wsol}
\eeq
where $\cE$ is the shell proper energy, we find up to $\cO(\hbar)$,
\bea
&&B S_0{'^2} + (m-\cE)^2 -\frac{f^2}B = 0\cr\cr
&&\sqrt{B}(\sqrt{B} S_0')' + \frac{2B}A A' S_0' = 0.
\eea
The first equation is solved by the Hamilton-Jacobi function,
\beq
S_0(R) = \pm \int \frac{dR}B \sqrt{f^2 - (m-\cE)^2B},
\label{S0}
\eeq
and the second gives
\beq
A = \frac C{|B|^{1/4}\sqrt{|S_0'|}}.
\eeq
Let us now show that the classical limit of this solution yields the classical dynamical equations that follow from \eqref{ptham}. To 
order $\hbar^0$, $W(\tau,R)$ is just the Hamilton-Jacobi function, so the function $R(\tau)$ defined by the principle of constructive 
interference,
\beq
\frac{\partial S_0}{\partial \cE} = 0 = -\tau \pm \int \frac{(m-\cE)dR }{\sqrt{f^2 - (m-\cE)^2B}}
\label{CI}
\eeq
and 
\beq
P(\tau) = S_0' = \pm\frac 1B\sqrt{f^2 - (m-\cE)^2B}
\label{hjP}
\eeq
should satisfy the Hamiltonian equations based on \eqref{ptham}. Taking a derivative of \eqref{CI},
\beq
1 = \pm \frac{R_\tau(m-\cE)}{\sqrt{f^2 - B(m-\cE)^2}}
\eeq
and therefore
\beq
m-\cE = \frac f{\sqrt{B+R_\tau^2}}
\eeq
Inserting this into \eqref{hjP} shows that
\beq
BP = \sqrt{f^2 - B(m-\cE)^2} = \frac{fR_\tau}{\sqrt{B+R_\tau^2}}
\eeq
or
\beq
R_\tau = \frac{BP}{\sqrt{\frac{f^2}B-BP^2}} = \left\{R,\cH\right\}
\label{RPB}
\eeq
Again, taking the derivative of $P$ in \eqref{hjP} and using \eqref{RPB} we find
\beq
P_\tau = \frac{2ff'-(f^2/B+BP^2) B'}{B\sqrt{\frac{f^2}B-BP^2}} = \left\{P,\cH\right\}.
\eeq
It follows that the trajectories implied by the principle of constructive interference in \eqref{CI} are identical to those 
determined by the Hamiltonian equations of motion that follow from \eqref{ptham}.

The WKB solutions may now be given as
\beq
\Psi_\pm(\tau,R) = \frac{C_\pm e^{-i\cE\tau}}{|B^{1/4}\sqrt{S_0'}} \exp \left[\pm i \int \frac{dR}B \sqrt{f^2-(m-\cE)^2B}\right].
\eeq
Inside the black hole horizon, because $B<0$, the the WKB wave function is always oscillatory, but the situation is different 
outside. The phase $S_0$ is real in the classically allowed region, for which $f^2 - (m-\cE)^2B>0$, and imaginary in the 
classically forbidden region. The wave function thus falls off exponentially when $f^2 - (m-\cE)^2 < 0$ {\it i.e.,} 
when
\beq
R> R_+ = \frac 12 \left(\kappa + \sqrt{\kappa^2 + \frac{\mu^4}{(m-\cE)^2}}\right),
\eeq
where $\kappa = 2GM^-$ and $\mu^2 = Gm^2 = (m/m_p)^2$. We will henceforth distinguish between the ``interior'' region ($R<\kappa$) 
and the ``exterior'' region ($R>\kappa$), which itself consists of the classically allowed region ($\kappa < R < R_+$) and the 
classically forbidden region ($R_+<R$). In the interior we write
\bea
\Psi(\tau,R) &=&\frac{e^{-i\cE\tau}}{|B|^{1/4}\sqrt{|S_0'|}}\left\{F_1 \exp\left[+i\int\frac{dR}B\sqrt{f^2 - (m-\cE)^2B}\right]\right.\cr\cr
&&\hskip 2cm \left. + F_2 \exp\left[-i\int \frac{dR}B\sqrt{f^2 - (m-\cE)^2B}\right]\right\},~~ 0 < R < \kappa
\eea
and in the exterior,
\beq
\Psi(\tau,R) = \left\{\begin{matrix}
\frac{e^{-i\cE\tau}}{|B|^{1/4}\sqrt{|S_0'|}} \left\{D_1 \exp\left[+i\int \frac{dR}B\sqrt{f^2 - (m-\cE)^2B}\right]\right.\cr\cr 
\left. \hskip 2cm + D_2 \exp\left[-i\int \frac{dR}B\sqrt{f^2 - (m-\cE)^2B}\right]\right\}, & \kappa < R<R_+\cr\cr
\frac{e^{-i\cE\tau}}{|B|^{1/4}\sqrt{|S_0'|}} \left\{D_3\exp\left[-\int \frac{dR}B\sqrt{(m-\cE)^2B-f^2}\right]\right.\cr\cr 
\left. \hskip 2cm + D_4 \exp\left[+\int\frac{dR}B\sqrt{(m-\cE)^2B-f^2}\right]\right\}, & R>R_+
\end{matrix}\right.
\eeq
where $D_j$ and $F_j$ are constants. The wave functions in the exterior {\it i.e.,} in the classically allowed and forbidden regions, 
can be matched in the standard way by invoking the asymptotic forms of the Airy functions far from the boundary between the regions. 
One readily finds the connection rules,
\beq
D_1 = \left(D_3 - \frac i2 D_4\right)e^{i\pi/4},~~ D_2 = \left(D_3 + \frac i2 D_4\right)e^{-i\pi/4}.
\eeq
Since the classically forbidden region extends to infinity we take $D_4 = 0$, which implies that $D_1 e^{-i\pi/4} = D_2 e^{i\pi/4} =  D_3$. 
In remains to match the interior and exterior solutions at the horizon, where the integral defining the phase has an essential singularity 
of order one.

We define the integral by analytically continuing to the complex plane, deforming the path so as to go around the pole at $R=\kappa$ in 
an infinitesimal circle of radius $\varepsilon$. Let $C_\varepsilon$ denote the deformed path, $S_\varepsilon$ the semi-circle of radius 
$\varepsilon$ about $R=\kappa$ in the complex $R$ plane, then we {\it define}
\beq
\int^R \frac{dR}{\sqrt{B}}\sqrt{\frac{f^2}B - (m-\cE)^2}~ \stackrel{\text{def}}{=}~ \lim_{\varepsilon\rightarrow 0} 
\underset{(C_\varepsilon)}{\int^R} \frac{dR}{\sqrt{B}}\sqrt{\frac{f^2}B - (m-\cE)^2}
\eeq
and choose the orientation of the semi-circle as a boundary condition. Performing the integration from left to right for 
$R=\kappa +\varepsilon$
\bea
\underset{(C_\varepsilon)}{\int^{\kappa+\varepsilon}} \frac{dR}B\sqrt{f^2 - (m-\cE)^2B} &=& \underset{(C_\varepsilon)}
{\int^{\kappa-\varepsilon}} \frac{dR}B\sqrt{f^2 - (m-\cE)^2B}\cr\cr 
&& \hskip 1cm + \underset{(S_\varepsilon)}{\int}\frac{dR}B\sqrt{f^2 - (m-\cE)^2B}
\eea
Since $\varepsilon$ is small, we perform a near horizon approximation of the integrand in the second integral,
\beq
\underset{(S_\varepsilon)}{\int}\frac{dR}B\sqrt{f^2 - (m-\cE)^2B} \approx \int_{S_\varepsilon}\frac{dR Rf}{(R-\kappa)} = 
\pm \frac{i\pi \mu^2}2 
\eeq
where the positive sign occurs if the path is deformed in the lower half complex plane, the negative sign occurs when the 
path is deformed in the upper half complex plane. In the present situation, there appears to be no good reason to choose one 
over the other. Each choice amounts to the identifications
\beq
F_1 = D_1 e^{\mp\pi\mu^2/2} = D_3 e^{i\pi/4}e^{\mp\pi\mu^2/2},~~  F_2 = D_2 e^{\pm\pi\mu^2/2} = D_3 e^{-i\pi/4} e^{\pm\pi\mu^2/2}.
\eeq
Owing to the sign change in $B$ across the horizon, outgoing waves in the exterior are matched to infalling waves in the interior 
and, vice-versa, infalling waves in the exterior are matched to outgoing waves in the interior, and the complete wave function is
\beq
\Psi(\tau,R) =\left\{\begin{matrix}
\frac{D_3e^{-i\cE\tau}}{|B|^{1/4}\sqrt{|S_0'|}}\left[e^{\mp\pi\mu^2/2} e^{\frac{i\pi}4}e^{iS_0}+ e^{\pm\pi\mu^2/2}e^{-\frac{i\pi}4}
e^{-iS_0}\right] & 0<R<\kappa\cr\cr
\frac{2D_3e^{-i\cE\tau}}{|B|^{1/4}\sqrt{|S_0'|}}\cos \left[S_0 + \frac \pi 4\right], & \kappa < R< R_+\cr\cr
\frac{D_3e^{-i\cE\tau}}{|B|^{1/4}\sqrt{|S_0'|}}e^{-\int |S_0'|dR}, & R>R_+ \end{matrix}\right. 
\eeq
where $S_0(R)$ is defined in \eqref{S0}.

\section{The Energy Spectrum}

For any two solutions of the wave equation in \eqref{wd} there is a conserved bilinear current density given by 
\beq
J_i = - \frac i2 \Phi^*\overleftrightarrow{\nabla}_i\Psi + m \delta_{i\tau}\Phi^*\Psi
\eeq
the time component of which determines a physical inner product 
\beq
\langle \Phi,\Psi\rangle = \int \frac{dR}{\sqrt{B}}\left[- \frac i2 \Phi^*\overleftrightarrow{\nabla}_\tau\Psi + m \Phi^*\Psi\right].
\eeq
Consider two stationary states,
\beq
\Psi_\cE(\tau,R) = e^{-i\cE\tau}\psi_\cE(R),~~ \Phi_{\cE'}(\tau,R) = e^{-i\cE'\tau}\psi_{\cE'}(R)
\eeq
of energies $\cE$ and $\cE'$, then
\beq
\langle \Phi,\Psi\rangle = \left[m-\frac 12(\cE+\cE')\right]e^{-i(\cE+\cE')\tau}\int \frac{dR}{\sqrt{B}}\phi^*_{\cE'}\psi_\cE
\label{inprodB}
\eeq
is positive semi-definite so long as $\cE<m$. By the wave equation we get 
\bean
\phi^*_{\cE'}\sqrt{B}\partial_R\sqrt{B}\partial_R \psi_\cE &=& \left[(m-\cE)^2 - \frac{f^2}B\right]\phi^*_{\cE'}\psi_\cE\cr\cr
\psi_\cE\sqrt{B}\partial_R\sqrt{B}\partial_R \phi^*_{\cE'} &=& \left[(m-\cE')^2 - \frac{f^2}B\right]\phi^*_{\cE'}\psi_\cE
\eea
Subtracting the second from the first,
\beq
\sqrt{B}\partial_R \left(\sqrt{B} \phi^*_{\cE'}\overleftrightarrow{\partial_R}\psi_\cE\right) = (\cE-\cE')\left(\cE+\cE' - 2m\right)
\phi^*_{\cE'}\psi_\cE
\eeq
showing that the inner product in \eqref{inprodB} is just a surface term which, mindful of the three regions, we give as
\beq
\langle \Phi,\Psi\rangle = -\frac i{(\cE-\cE')}\left[\int_0^\kappa dR \partial_R \sqrt{B}J_R + \int_\kappa^{R_+} dR \partial_R 
\sqrt{B}J_R + \int_{R_+}^\infty dR \partial_R \sqrt{B}J_R\right].
\eeq
To guarantee orthonormality of the wave functions, we must require that the inner product vanishes whenever $\cE'\neq \cE$. 
Therefore, calling $\Omega = \sqrt{B}J_R$, we seek the conditions under which
\beq
\lim_{R\rightarrow \infty}\Omega - \lim_{R\rightarrow R_+^+}\Omega + \lim_{R\rightarrow R_+^-}\Omega - \lim_{R\rightarrow \kappa^+}
\Omega + \lim_{R\rightarrow \kappa^-}\Omega - \lim_{R\rightarrow 0}\Omega = 0,
\eeq
where the superscripts indicate the left/right limits. The first term vanishes because the wave function vanishes exponentially at 
infinity. Direct computation also shows that the second and third terms cancel and the last term vanishes but the fourth 
and fifth terms, which must be evaluated at the black hole horizon, neither separately vanish nor cancel one another. This 
occurs because the black hole horizon is an essential singularity of the phase integral. One finds
\beq
- \lim_{R\rightarrow \kappa^+} \Omega + \lim_{R\rightarrow \kappa^-}\Omega \sim \sin\left[\frac{\mu^2}2 \ln\left(\frac{m-\cE}{m-\cE'}
\right)\right],
\eeq
and therefore
\beq
\frac{m-\cE}{m-\cE'} =  e^{\frac{2n\pi}{\mu^2}}
\eeq
for integer values of $n$. Assuming the existence of a ground state, it implies the energy spectrum,
\beq
\cE_n = m\left(1-e^{-2n\pi/\mu^2}\right),
\eeq
where $n$ is a whole number. This is identical to the spectrum of the shell collapsing in a vacuum when $\mu\gg 1$, {\it i.e.,} when 
the proper mass of the shell is much larger than the Planck mass. The presence of the external black hole does not disturb the 
spectrum to this order.

\section{Conclusions}

The purpose of this work was to understand the similarities and differences in the quantum mechanics of a single thin shell and the midisuperspace 
quantization of the shells in a collapsing dust ball. We examined the WKB approximation to the proper time quantum mechanics of the thin 
dust shell when it surrounds a pre-existing black hole. We have shown that although the construction of the Hamiltonians governing the evolution 
of the two systems have very different origins (the dynamics of the thin shell are obtained via an application of the IDL conditions whereas the 
dynamics of the dust ball are fully derived from the Einstein-dust system) the Hamiltonians one ends up are structurally similar with a crucial 
exception: the thin shell posseses a self-interaction that depends on its area radius, but the shells of a dust ball do not. This self interaction 
causes the thin shell to always stay bound to the center, regardless of whether the interior of the shell is a vacuum or a black hole. On the 
contrary, the shells of a dust ball may be unbound. Again as a consequence of the self-interaction, the matching of the wave function at the 
horizon of the black hole is accomplished with essentially no information about the horizon length, only the proper mass of the shell plays a role. 
Neither does the black hole play any role in the energy spectrum of the shell, which we have shown is identical to the spectrum 
of a shell collapsing in a vacuum to this order.

What is most surprising is that bound state solutions exist for shell proper masses greater than the Planck mass and the shell does {\it not} 
collapse into the central singularity, suggesting that quantum uncertainty plays a role in the collapse over distance scales determined 
by the size of the black hole's event horizon and larger than previously suspected. The wave function vanishes at the center, extends out to the 
turning point, which, depending on the energy of the shell, may lie close to the horizon but always outside it, and falls off exponentially 
beyond this point. This is reminiscent of a gravitational atom and supports another solution of the quantum dust ball collapse, which does not 
involve continued collapse as described in the introduction: if the collapse does not continue past the apparent horizon, the solution is 
described by the first of the two solutions given in the introduction and the shells will coalesce on the apparent horizon. No event horizon 
will form and the collapse will end in an ultra compact star instead of a black hole \cite{hawking14,vaz15}. 

In general, proper time quantization seems to enjoy several advantages over coordinate time quantizations. For one, the proper time quantum theory 
exists for shells of mass greater than the Planck mass, unlike the quantum mechanics that is based on coordinate time. But the most important is 
that it satisfies a basic requirement of the quantum theory, {\it i.e.,} observer independence of the time parameter. It is therefore ``democratic'' 
in regard to all foliations of spacetime: all coordinate time variables would be functions of the phase space (in the simple case of the shell these 
are given by \eqref{timeders}) as are the spatial coordinates.  The same would be true of the metric components. Thus they would all be operator 
valued and we would be able to speak of time intervals and spatial distances only in terms of averages. In the proper time formulation, these averages 
can be calculated and fluctuations about them quantified because the Wheeler-DeWitt equation yields a conserved, positive, semi-definite inner product.

\end{document}